\documentclass{article}

\usepackage[final]{neurips_2022}
\usepackage{enumitem}

\usepackage[small,compact]{titlesec}
\usepackage[utf8]{inputenc} % allow utf-8 input
\usepackage[T1]{fontenc}    % use 8-bit T1 fonts
\usepackage{hyperref}       % hyperlinks
\usepackage{url}            % simple URL typesetting
\usepackage{booktabs}       % professional-quality tables
\usepackage{amsmath}
\usepackage{amsfonts}       % blackboard math symbols
\usepackage{nicefrac}       % compact symbols for 1/2, etc.
\usepackage{microtype}      % microtypography
\usepackage{xcolor}         % colors
\usepackage{subcaption}
\usepackage{graphicx}
\usepackage{caption}
\usepackage{comment}
\usepackage[scaled=.8]{beramono}
\newcommand{\blue}[1]{\textcolor{black}{#1}}
\newcommand{\Fig}[1]{Figure~\ref{#1}}

\title{Multi-Agent Reinforcement Learning for Microprocessor Design: \\
A DRAM Memory Controller Design Case Study}
\title{Multi-Agent Reinforcement Learning for Microprocessor Design Space Exploration: \\A Case Study on DRAM Memory Controller Design}
\title{Multi-Agent Reinforcement Learning for Microprocessor Design Space Exploration} \vspace{-10pt}
\author{%
  Srivatsan Krishnan\thanks{Work done, in part, via the Google Student Researcher program. Contact at srivatsan@seas.harvard.edu}\\ Harvard University \\ \& \\ Google Research \\ Brain Team\And
   Natasha Jaques\\ Google Research\\Brain Team\And 
   Shayegan Omidshafiei\\ Google Research\\People + AI Research Team \And
   Dan Zhang\\ Google Research\\Brain Team\AND
   Izzeddin Gur\\ Google Research\\Brain Team\And
  Vijay Janapa Reddi\\ Harvard\\University\And
  Aleksandra Faust\\ Google Research\\Brain Team\And
}

\begin{document}

\maketitle

\begin{abstract}
Microprocessor architects are increasingly resorting to domain-specific customization in the quest for high-performance and energy-efficiency.
%Architectural design space exploration (DSE) is a fundamental exploration task that architects perform in the quest for finding efficient hardware designs. 
As the systems grow in complexity, fine-tuning architectural parameters across multiple sub-systems (e.g., datapath, memory blocks in different hierarchies, interconnects, compiler optimization, etc.) quickly results in a combinatorial explosion of design space. This makes domain-specific customization an extremely challenging task. Prior work explores using reinforcement learning (RL) and other optimization methods to automatically explore the large design space. However, these methods have traditionally relied on single-agent RL/ML formulations. It is unclear how scalable single-agent formulations are as we increase the complexity of the design space (e.g., full stack System-on-Chip design). Therefore, we propose an alternative formulation that leverages Multi-Agent RL (MARL) to tackle this problem. The key idea behind using MARL is an observation that parameters across different sub-systems are more or less independent, thus allowing a decentralized role assigned to each agent. We test this hypothesis by designing domain-specific DRAM memory controller for several workload traces. Our evaluation shows that the MARL formulation consistently outperforms single-agent RL baselines such as Proximal Policy Optimization and Soft Actor-Critic over different target objectives such as low power and latency. To this end, this work opens the pathway for new and promising research in MARL solutions for hardware architecture search.
\end{abstract}

\section{Introduction}

%The demand for high-performance and energy-efficient computer systems continues to rise despite the impending doom of Moore's law. As a result, system architects continue to rely on domain-specific specialization to meet the stringent latency, power, and energy requirements. These domain-specific specializations are tailor-made hardware resource selections for a given application with better performance/energy profiles than a one-size fits all general-purpose computer system. 

Computer systems are complex and interconnected systems, where various sub-components interact with each other to execute a program successfully. Due to the slowdown of Moore’s law~\citep{moore} and end of Dennard scaling~\citep{dennard}, computer system architects are increasingly resorting to domain-specific customization~\citep{dsa} of various sub-components in the computer system (e.g., compute, cache, memory, interconnects) to achieve better performance under stringent energy efficiency targets. As the customization boundary extends to the whole computer system stack (i.e., datapath, memory, interconnects, compilers, and application), it also results in a combinatorial explosion in the number of possible custom designs. For instance, enumerating the full stack for a custom DNN accelerator increases the overall parameter space to 10$^{2300}$~\citep{dan_asplos}.

%Prior work~\citep{dsa} has shown that domain-specific architecture specialization significantly improves performance and energy metrics for a wide variety of workloads. However, the downside of pursuing hardware specialization for meeting aggressive performance/energy targets is the combinatorial explosion in the design space. To make matters worse, exposing all the parameters across the whole system results in an extremely large design space that becomes intractable to search the space manually. For instance, the size of the design space for full-system cross stack optimization (e.g., parameters from datapath, memory blocks in different hierarchies, interconnects, compiler optimization etc) for DNN accelerators results in the order of 10$^{2300}$~\citep{dan_asplos}.

%To navigate this large design space intelligently, system architects rely on single-agent ML/RL formulations to quickly navigate the large design space. However, as more hardware parameters are exposed, it is unclear if the current single-agent formulations will scale to keep up with the complexity of the computer systems. 

\textbf{Related Work and Current Approach.} The popularity~\citep{rl-book} of reinforcement learning (RL) continues to rise, and computer architects have successfully applied RL for the parameter selection problem. For instance, RL-based parameter selection was applied for designing a custom scheduling policy in memory controllers~\citep{rl_mem_ctrl}, data path for DNN accelerators~\citep{rl_data_path,rl_data_path_2}, Network-on-Chip (NoC) topology~\citep{rl_nocs}, and transistor sizing~\citep{rl_transistor, rl_transistor_2}. However, these RL formulations are single-agent based, where a single policy simultaneously determines all system parameter values. In these scenarios, the policy needs to learn the observation to action mapping and the relationship between various parameters in all components of the interconnected system. As the complexity of the computer system increases, it is unclear if single-agent formulations will scale accordingly.  

\begin{figure*}[t!]
  \centering
% \begin{subfigure}[t]{0.3\columnwidth}
%   \centering
%   \includegraphics[width=0.75\columnwidth]{figs/interconnected_components.png}
%   \caption{Interconnected Systems.}
%   \label{fig:interconnected_systems}
% \end{subfigure}%
\begin{subfigure}[t]{0.55\columnwidth}
  \centering
  \raisebox{20pt}{\includegraphics[width=\columnwidth]{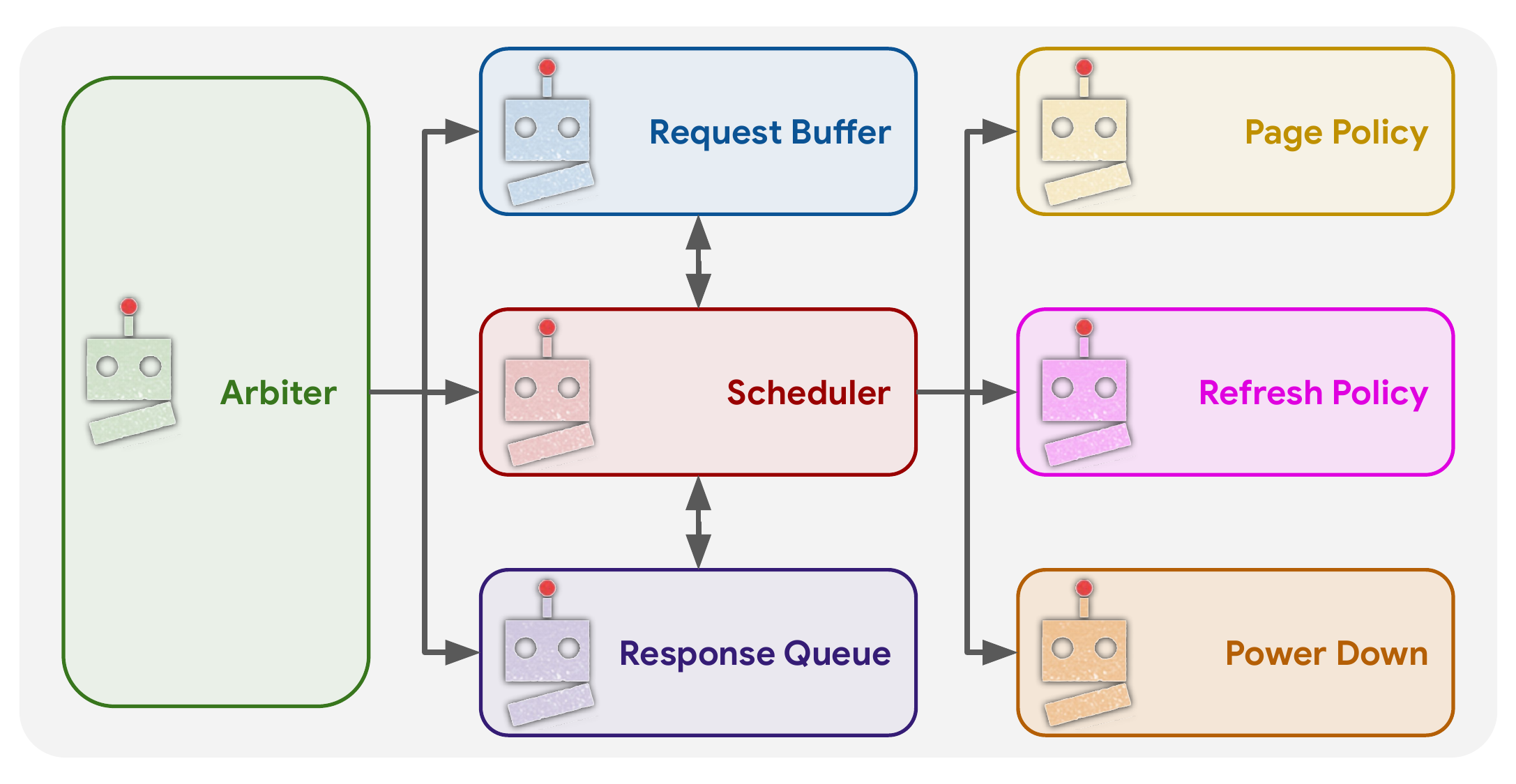}}
  \caption{Interconnected components and MARL formulation.}
  \label{fig:interconnected_systems_marl}
\end{subfigure}%
\begin{subfigure}[t]{0.4\columnwidth}%
  \centering
  \includegraphics[width=0.9\columnwidth]{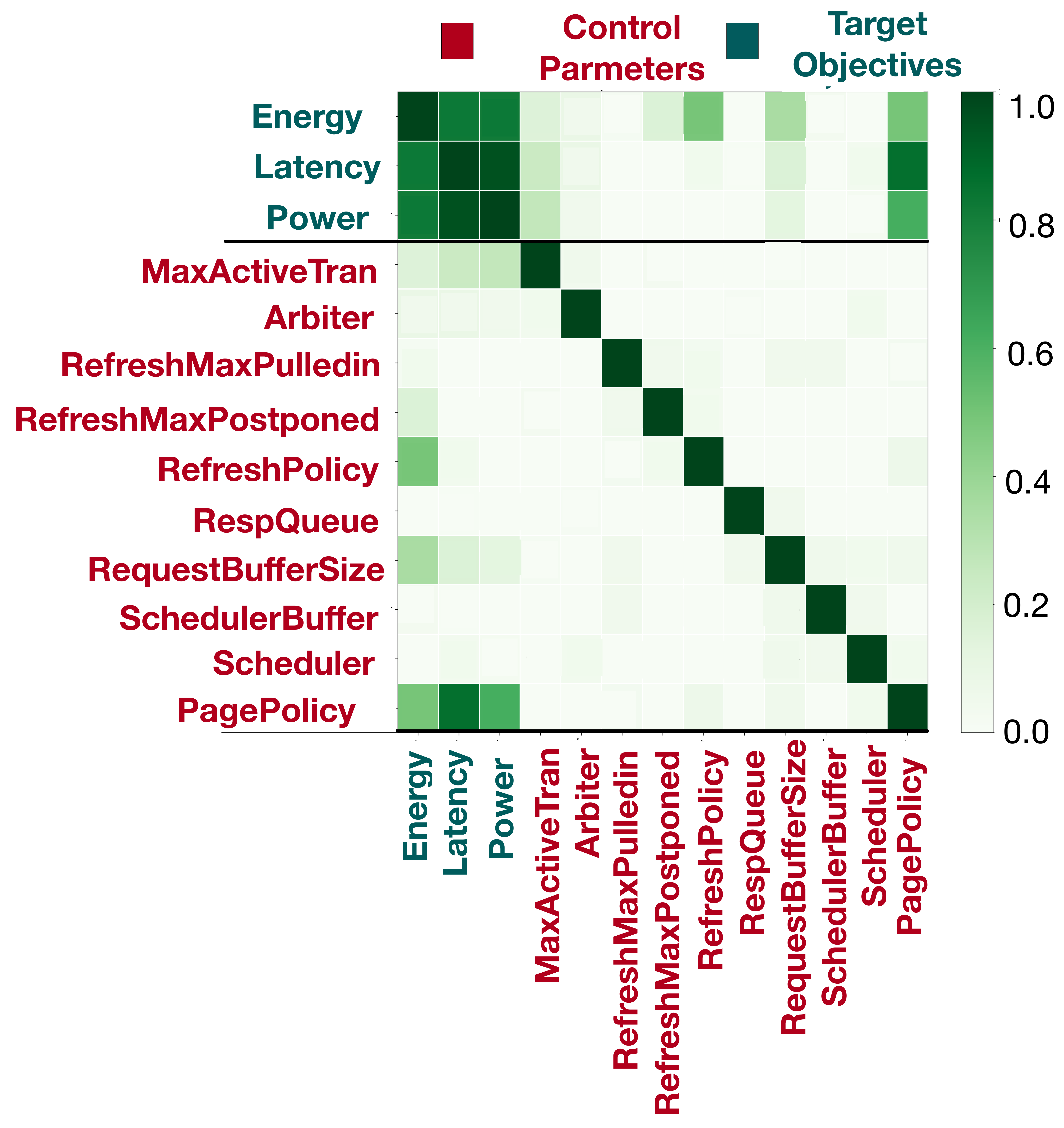}
  \caption{Correlation matrix.}
  \label{fig:corr_mat}
\end{subfigure}%
% \begin{subfigure}[t]{0.34\columnwidth}
%   \centering
%   \includegraphics[width=0.95\columnwidth]{figs/marl_interconnected_agent.png}
%   \caption{Multi-Agent Formulations.}
%   \label{fig:multi_agent}
% \end{subfigure}
% \caption{ (a) DRAM Memory controller as an example of complex interconnected system. (b) $\phi_k$ correlation~\citep{phik} between various memory controller parameters and target objectives. (c) MARL formulation for DRAM memory controller design. }
\caption{ (a) DRAM memory controller as an example of an interconnected system. Also indicated is the proposed MARL formulation, involving each agent controlling a distinct system component. (b) $\phi_k$ correlation~\citep{phik} between  memory controller parameters and target objectives. 
\vspace{-1.5em}}

%\caption{DRAM memory controller parameter exploration using different RL formulations. We only show four parameters in the memory controller for brevity. The parameters from core and DRAM chips are not the focus and hence faded out. (a) Single Agent RL formulation (b) Multi-Agent RL formulation for architecture parameter search problems.\vspace{-20pt}}

\end{figure*}

\textbf{Our Contribution.} We introduce Multi-Agent RL (MARL) to tackle the problem of microprocessor hardware architecture search in a scalable and systematic manner. To the best of our knowledge, this is the first multi-agent formulation to tackle architecture design space exploration. To apply our MARL formulation, we use a DRAM memory controller as an example of an interconnected system shown in \Fig{fig:interconnected_systems_marl}. A DRAM memory controller consists of \texttt{Request Buffer}, \texttt{Response Buffer}, \texttt{Arbiter}, \texttt{Scheduler}, and policy control managers for determining \texttt{Page Policy}, \texttt{Refresh Policy}, and \texttt{PowerDown Policy}. Each of these subsystems can have many parameters which eventually determine the latency and power consumption for the DRAM subsystem. Thus, the DRAM memory controller plays a key role in determining the performance of the memory subsystem. %Prior work has shown the performance of the memory system plays a significant role in the computer system's overall performance. 

\textbf{\blue{Case for }Decentralized MARL.} The key idea behind using a decentralized MARL approach is that the parameters in a computer system are more or less independent (are not strongly correlated with each other). \blue{Intuitively, for high-performance and energy-efficient designs, we want each component to be independent since having more than one sub-components achieve the same purpose means having redundancy in the system. Though designing a redundant system is a valid design goal, our focus in this work is to design customized hardware architectures for high performance and efficiency.}  

%\blue{The observation of having independent components} can be leveraged to create a system of decentralized agents, each controlling one parameter in an interconnected computer system. 
\blue{To empirically validate that the control parameters are independent, we perform a random grid search over the DRAM memory controller parameters and their impact on overall latency, energy, and power. Using this data, we construct the correlation matrix as shown in} \Fig{fig:corr_mat} between various parameters in a DRAM memory controller system. We observe that the control parameters have a very low correlation between them (darker shades of green mean more correlation) which makes it amenable to formulating the problem using a decentralized multi-agent approach. \blue{Note that this also motivates using decentralized multi-agent PPO~\citep{marl_ppo_1}  for architecture DSE, rather than centralized training, decentralized execution multi-agent RL formulations such as MAPPO~\citep{marl_ppo_2} or MADDPG~\citep{maddpg}}.

A key benefit of using MARL, rather than single-agent RL, is to divide and conquer large parameter design spaces. The MARL formulation provides the necessary constructs and infrastructure to assign fine-grained roles to many agents. Each agent implicitly communicates with other agents through shared rewards and observations to converge on a set of parameters such that the team meets the global design objective (e.g., low latency or low power).

In our MARL formulation for the DRAM memory controller, we assign each agent to control one specific memory controller parameter. For example, as shown in \Fig{fig:interconnected_systems_marl}, the \textit{Request Buffer Agent} controls the \texttt{Request Buffer} size. Similarly, the \textit{Page Policy Agent} controls the \texttt{Page Policy} parameter. Since the parameter space is divided across each independent agent, the MARL formulation allows for factorizing the large design space across many agents.

%We take DRAM memory controller design to show the benefits of our MARL formulation. Memory system performance plays a huge role in the computer system's overall performance. Though several prior works have heavily focused on the customization of computing stack (e.g., hardware accelerators), designing domain-specific memory systems can also play an important role in improving the performance and energy of the computer system.

\textbf{Domain-Specific Custom Memory Controller Design.} To validate the effectiveness of our MARL formulation, we design a custom memory controller for two different memory access patterns for lowering its latency and power.%, and a combination of lower latency and power.
For estimating the cost in terms of latency, power, and energy for selecting specific memory controller parameters, we use DRAMSys~\citep{dramsys}. DRAMSys is a transaction-level simulator that takes memory traces as inputs and outputs the latency, power, and energy for a given memory controller parameters. We construct a DRAM-Gym environment to evaluate our MARL formulation and variants of single-agent RL algorithms such as Proximal Policy Optimization (PPO)~\citep{ppo} and Soft Actor-Critic (SAC)~\citep{sac}. PPO and SAC are state-of-the-art RL algorithms that have shown promising results in several sequential decision-making tasks. We use the ACME~\citep{acme_2022} as the RL framework for the MARL and other single-agent baselines.% Furthermore, an exhaustive empirical study shows that the MARL approach consistently outperforms single-agent RL formulation across different optimization objectives and workload traces.

\textbf{Contributions.} We make the following contributions: (1) First demonstration of applying MARL for the microprocessor hardware architecture search problem, (2) We demonstrate our MARL formulation consistently outperforms single agent RL formulations in terms of designing a domain-specific memory controller for multiple memory traces and objective targets, (3) Our MARL formulations achieve a mean episodic reward that is $\approx$ 2$\times$ to 60$\times$ more than single-agent formulations.

\section{Methodology}
\label{sec:methodolgy}
%This section describes our experimental methodology for evaluating single-agent RL and multi-agent RL formulation for designing custom memory controllers for different workloads traces. 
% We first describe our MARL formulation and the different RL baselines used in this work. Next, we discuss the workloads for which we want to design a domain-specific memory controller. We then describe the DRAM-Gym environment, which encapsulates a DRAM simulator to evaluate our MARL formulation and other RL baselines. %We then discuss our MARL approach along with single-agent RL formulations as baselines.
% \textbf{RL Agents.}
In this work, we propose a MARL formulation for DRAM memory controller parameter exploration. We compare the performance of MARL formulations with single agent formulations such as PPO~\citep{ppo} and SAC~\citep{sac}. In particular, for the MARL formulation, we use multi-agent decentralized PPO since it has shown state-of-the-art performance in various sequential decision-making tasks~\citep{marl_ppo_1, marl_ppo_2}.

\paragraph{Decentralized MARL.} We have decentralized agents taking simultaneous independent actions. For the domain-specific custom memory controller design, we have ten parameters (see Appendix~\ref{app:params_mc}), each controlled by a distinct agent. For instance, the size of the \texttt{Request Buffer} is controlled by Request Buffer Agent whereas \texttt{Page Policy} is controlled by Page Policy Agent, as shown in \Fig{fig:interconnected_systems_marl}. Each agent receives the same observation and a shared reward signal. Please refer to Appendix~\ref{sec:rl-hyper} for more information.

\paragraph{Single-Agent Baselines.} For single-agent, we have three different baselines, namely PPO~\citep{ppo}, SAC~\citep{sac}, and what we define as Time Division Multiplexing (TDM). We use the ACME~\citep{acme_2022} framework for PPO and SAC, wherein a single agent takes ten actions simultaneously (see Appendix~\ref{app:params_mc}). Hyperparameters for these agents are tabulated in Appendix~\ref{sec:rl-hyper}. 

TDM is the most basic form of RL formulation~\citep{rl-book}, where a single agent produces one action each time step. In the case of the DRAM memory controller, there are ten parameters (refer to Appendix~\ref{app:params_mc}); thus, the TDM agent will predict \texttt{Page Policy} parameter at timestep one, the \texttt{Request Buffer Size} parameter at timestep two, and so on. At the end of ten timesteps, all the parameters accumulated are sent to the DRAM-Gym environment. Since there is a single policy controlled by a single agent (whose role changes on each timestep), we denote this formulation Time Division Multiplexing. Note that in the TDM formulation, the agent receives zero reward in intermediate timesteps; it only receives a non-zero reward signal at the end of all ten timesteps, once all parameter outputs are accumulated.
% while the agent performs TDM steps, it receives a zero reward. The TDM agent receives a non-zero reward once all the actions are taken. 
For more information on our TDM formulation, please refer to Appendix~\ref{sec:rl-hyper}.
% TDM RL formulations for the same optimization tasks. We refer the readers to Appendix~\ref{sec:rl-hyper} for more details about key hyperparameters.

%In our proposed MARL approach, we assign a fine-grained role to each agent where each parameter in Table~\ref{tab:mc_params} is controlled by one agent. The observation space and global reward signal are shared across all the agents. In contrast to MARL, in single-agent RL formulations, a single agent controls all the DRAM memory controller parameters. To remove statistical noise, we run a wide hyperparameter sweep for each evaluation involving multiple seeds, learning rate, and entropy values. We evaluate that trained policy at an interval of 100 training steps and report the mean episode return.

\textbf{Workload and Optimization Objectives.} From a DRAM perspective, a workload is a set of memory read/write accesses. Thus, we pick two extreme access patterns in this context: a streaming memory access trace and a random memory access trace. These are representative of real-world workloads. For instance, dense matrix-matrix multiplication and convolution operations would result in streaming access. Likewise, operations like pointer chasing and sparse matrix-matrix multiplication will result in random memory access. For more details, we refer the readers to Appendix~\ref{app:workload}.

% \textbf{Optimization Objective.} 
Our goal is to design a memory controller for three different objectives -- (a) Low latency, (b) Low Power, (c) Low latency and low power. %In this work, we target two memory traces: streaming address and random address memory trace. 
These are realistic objectives depending on the target domain. For instance, in an embedded use case, achieving low power might be critical, whereas achieving low latency in a high-performance computing domain is critical. We want to compare the performance of single-agent RL with multi-agent RL formulations against these optimization objectives.

\textbf{DRAM-Gym Environment.} 
We focus our empirical study on DRAMSys~\citep{dramsys}, which is a fast transaction-level simulator for modeling the performance of the DRAM memory system. DRAMSys takes memory traces as input and exposes several parameters in the memory subsystem, such as address mapping, memory controller parameters, and DRAM technology. Then, the simulator outputs the latency and power for that memory traces for a given parameter selection in the DRAM system.% To facilitate ease of integrating different RL formulations with DRAMSys, we construct DRAM-Gym, which provides an OpenAI gym interface for different RL algorithms. 
To evaluate the effectiveness of different RL methods, including multi-agent RL, we encapsulate DRAMsys into a Gym environment~\citep{brockman2016openai}, which provides a common interface for RL and MARL algorithms. 
% A Gym environment is an RL interface that exposes functions for \texttt{step}, \texttt{observation}, \texttt{reward}, and \texttt{reset} which makes it easy to integrate with standard RL algorithm implementations. 
We next describe the components of the RL formulation used in our experiments.
% To determine how latency and power vary as we change the memory controller parameters, we use DRAMSys. 

\textit{Observations.} In this version of the DRAM-Gym environment, the observation is a tuple of \texttt{<latency, power, energy>}. This observation is similar to a multigrid world environment where the objective is to reach a specific coordinate. In our case, the coordinates are an n-dimensional parameter space where each combination of the parameter space will have a corresponding \texttt{<latency, power, energy>}. We aim to find the specific coordinates in the n-dimensional space (i.e., memory controller parameters) that satisfy the optimization objectives.

\textit{Actions.} The considered action space corresponds to the parameters in the memory controller. The action space is diverse; some actions are categorical variables (e.g., \texttt{Page policy}), and some are integers (e.g., \texttt{Request Buffer Size}). For more information on the parameter space for the memory controller we explore in this work, we refer the readers to Appendix~\ref{app:params_mc} (Table~\ref{tab:mc_params}).

\textit{Rewards.}  We use similar reward structure for the objective of minimizing power and latency. For minimizing both, we use the following reward function
\begin{equation}
\centering
r_{x} = \frac{X_{\text{target}}}{|X_{\text{target}}-X_{\text{obs}}|} \, , \label{eq:r_power}
\end{equation}
where $X_{\text{target}}$ is the target power/latency and $X_{\text{obs}}$ is the currently-achieved value.
For the joint objective of minimizing both power and latency, we take the product of $r_{\text{joint}} = r_{\text{power}}*r_{\text{latency}}$.

\begin{figure*}[t!]
  %\centering
\begin{subfigure}[t]{0.5\columnwidth}
  %\centering
  \includegraphics[width=\columnwidth]{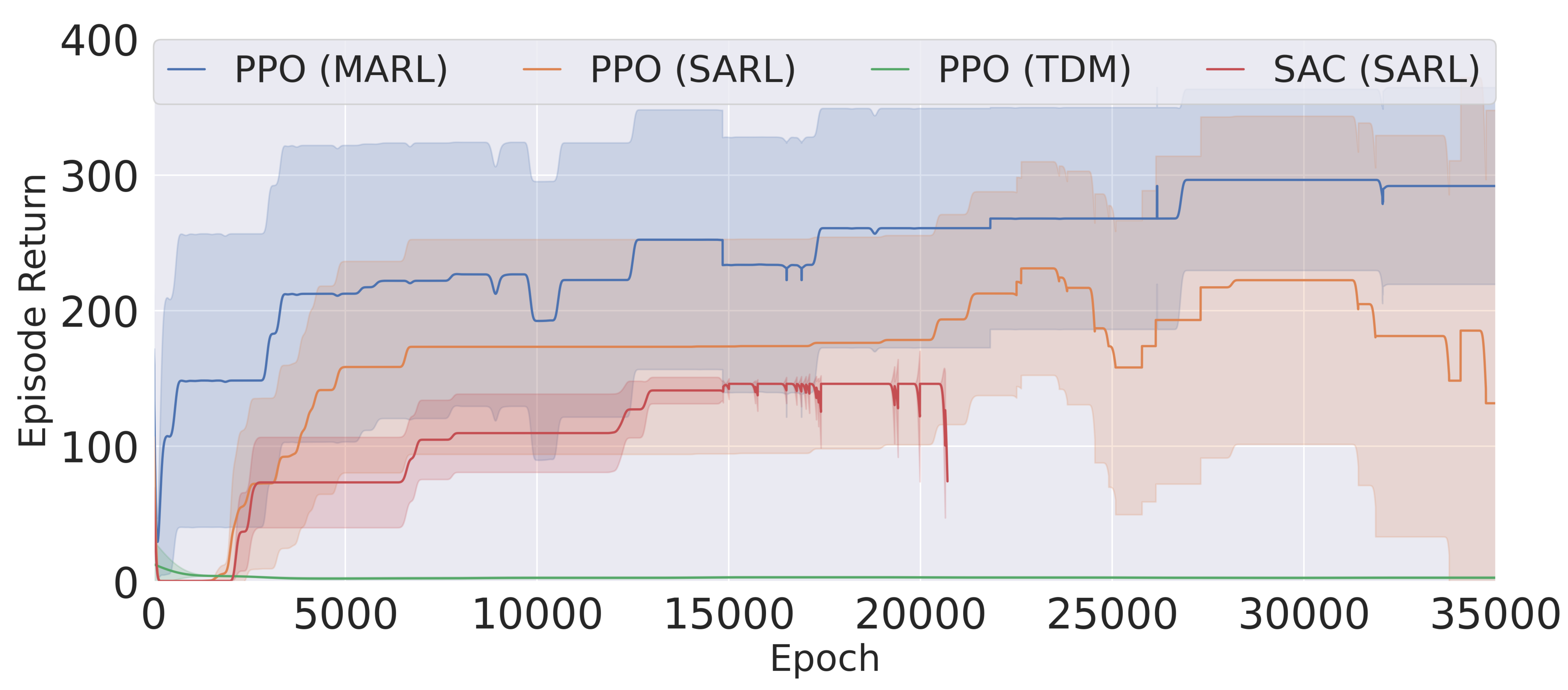}
  \caption{Low power. Random memory access trace.}
  \label{fig:rand-low-pow}
\end{subfigure}%
\begin{subfigure}[t]{0.5\columnwidth}
  %\centering
  \includegraphics[width=\columnwidth]{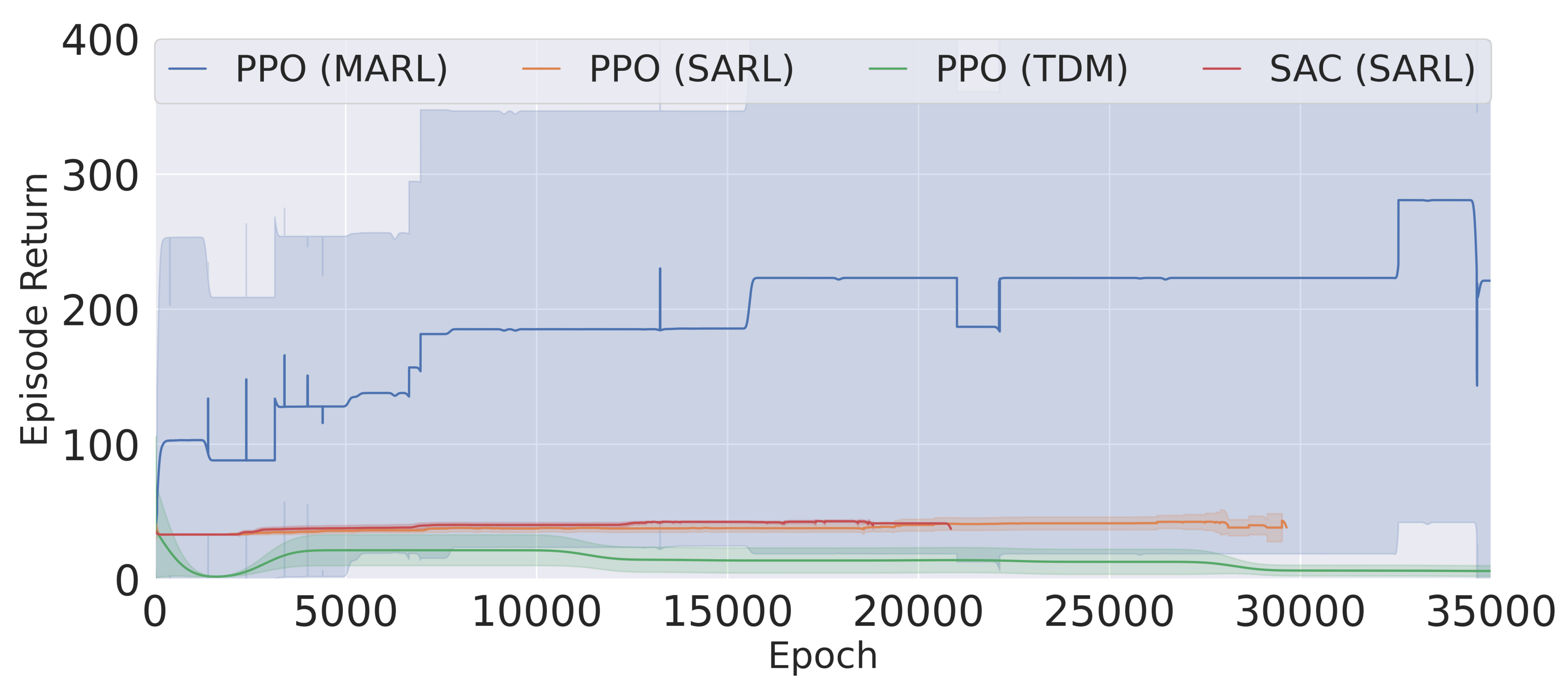}
  \caption{Low power. Streaming memory access trace.}
  \label{fig:stream-low-pow}
\end{subfigure}

\begin{subfigure}[t]{0.5\columnwidth}
  %\centering
  \includegraphics[width=\columnwidth]{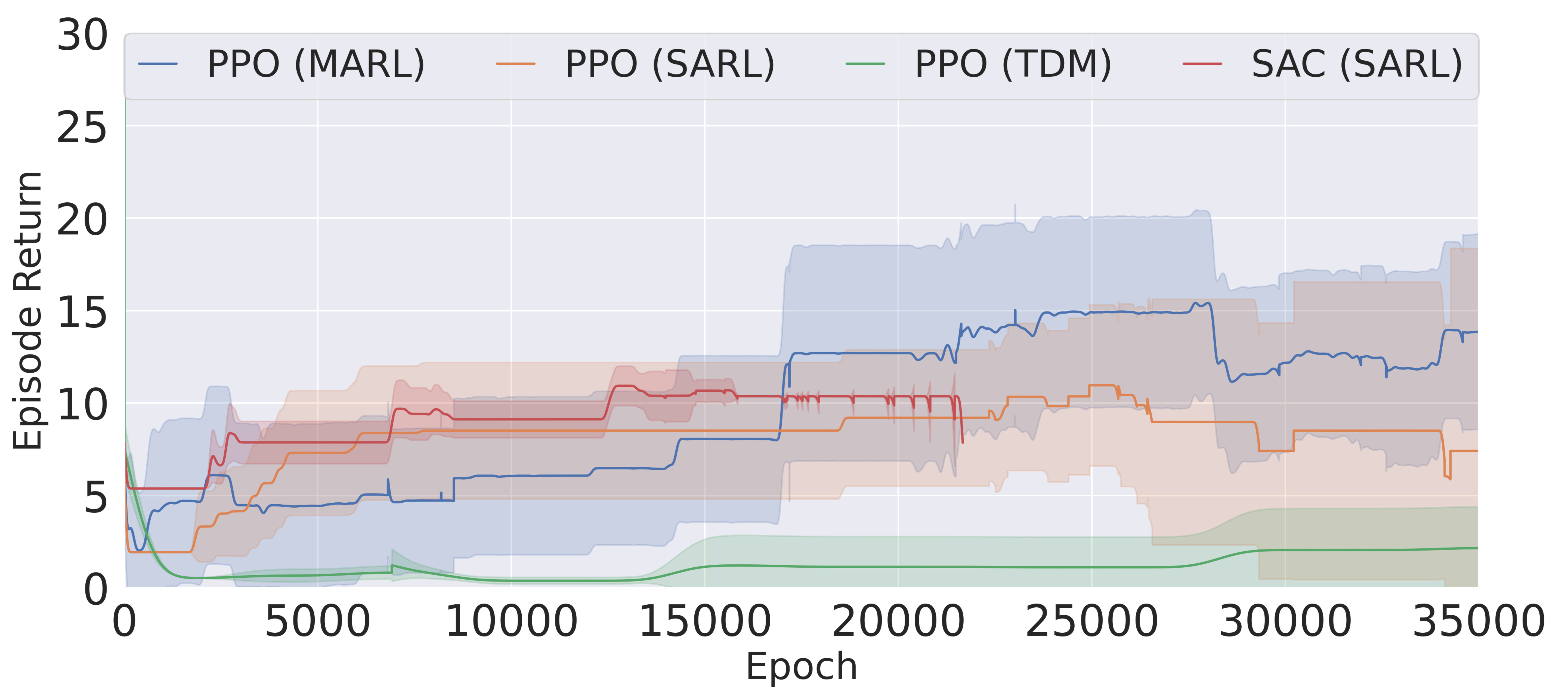}
  \caption{Low latency. Random memory access trace.}
  \label{fig:rand-low-lat}
\end{subfigure}%
\begin{subfigure}[t]{0.5\columnwidth}
  %\centering
  \includegraphics[width=\columnwidth]{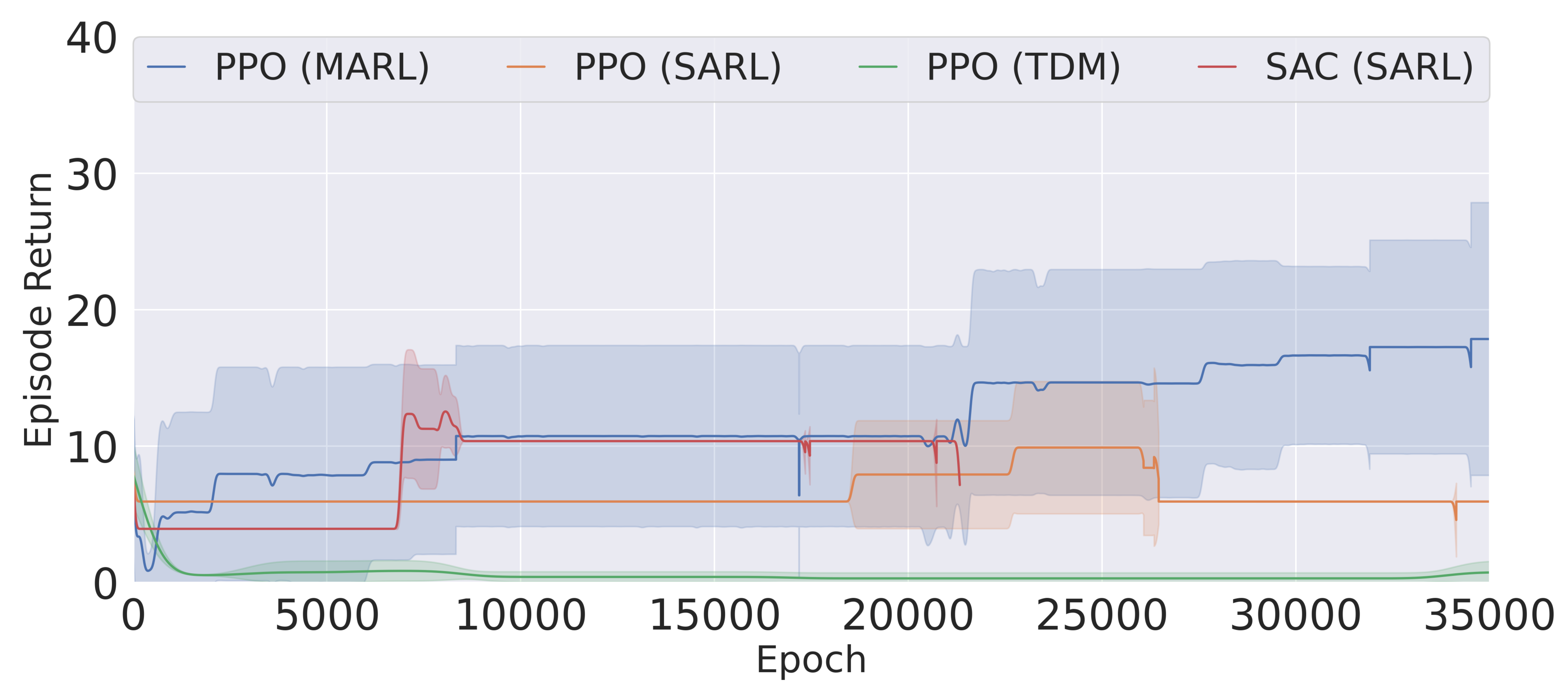}
  \caption{Low latency. Streaming memory access trace.}
  \label{fig:stream-low-lat}
\end{subfigure}

\begin{subfigure}[t]{0.5\columnwidth}
  %\centering
  \includegraphics[width=\columnwidth]{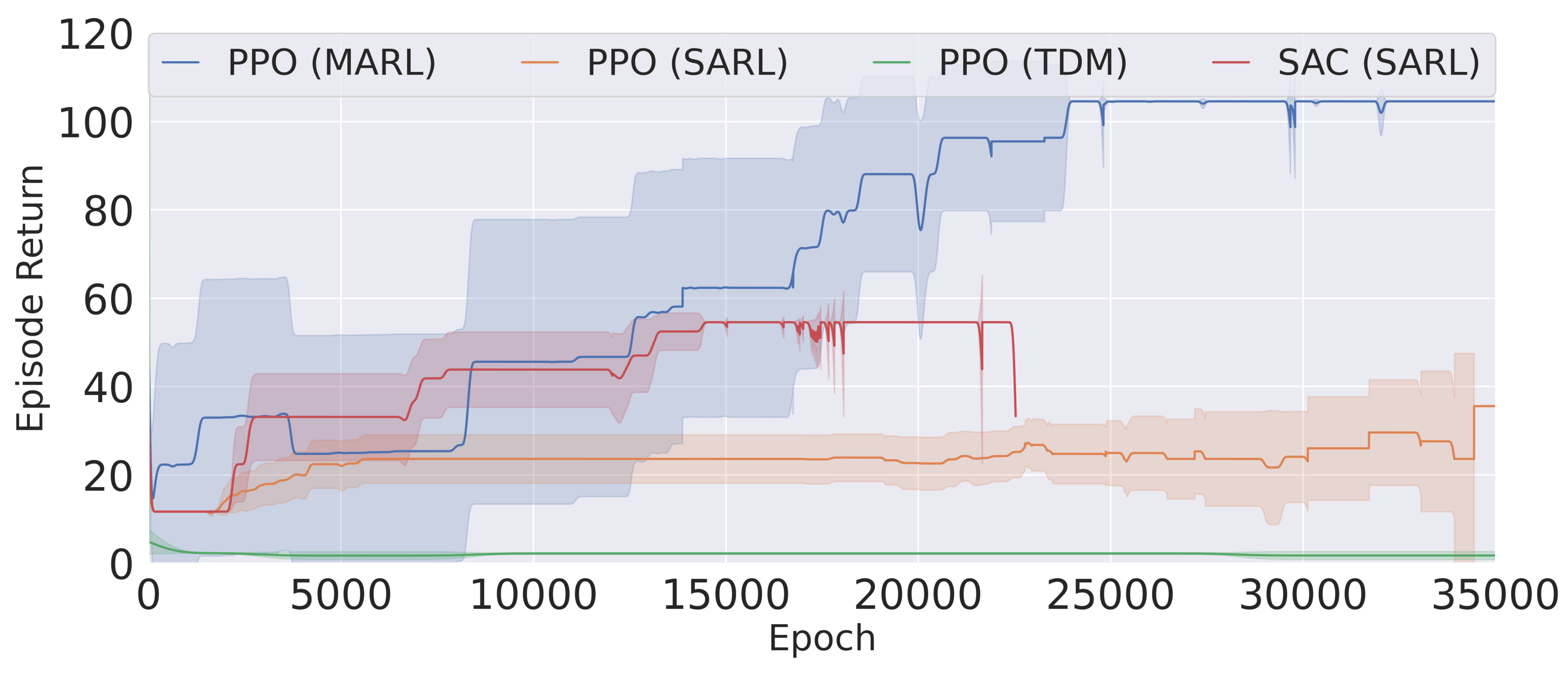}
  \caption{Joint optimization of random memory access.}
  \label{fig:rand-low-pow-lat}
\end{subfigure}%
\begin{subfigure}[t]{0.5\columnwidth}
  %\centering
  \includegraphics[width=\columnwidth]{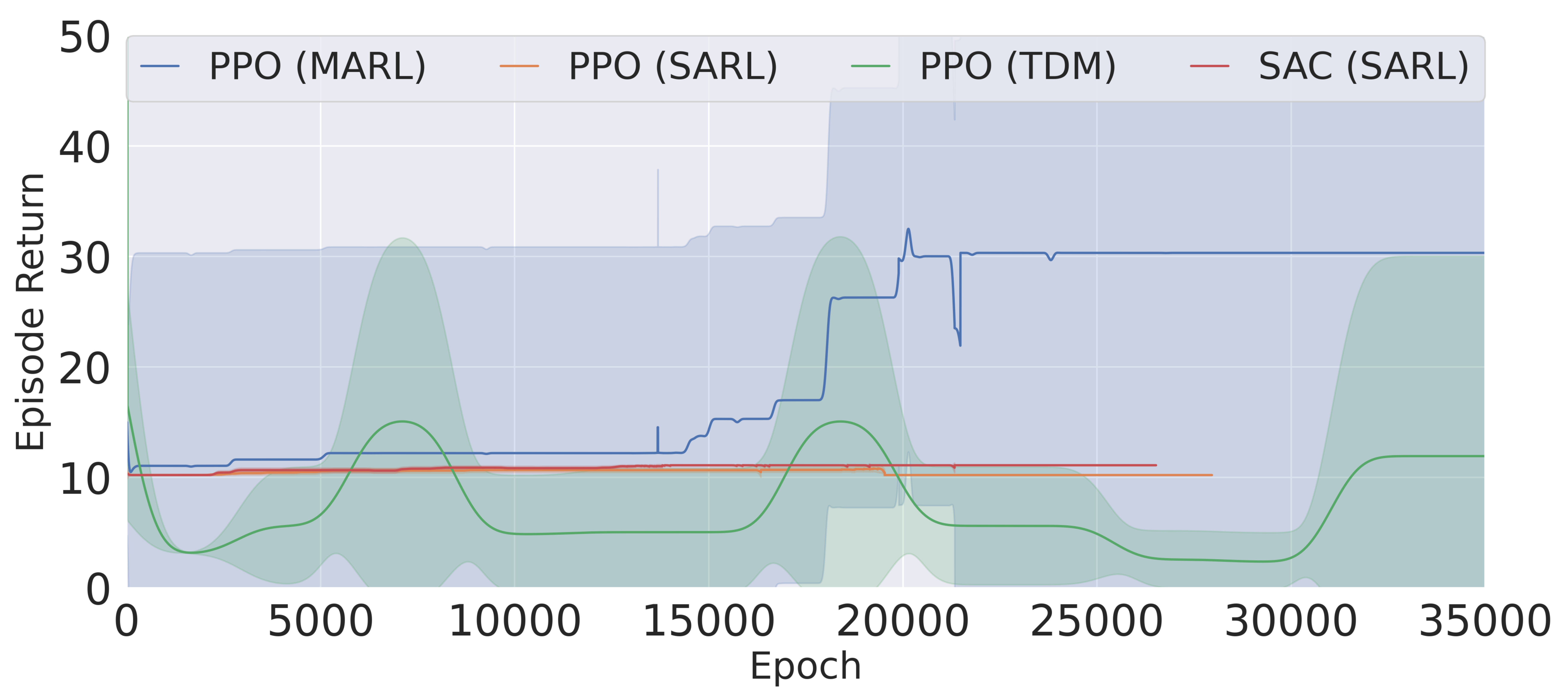}
  \caption{Joint optimization of streaming memory access.}
  \label{fig:stream-low-pow-lat}
\end{subfigure}%

\caption{(a),(c), and (e) Mean episode return for \texttt{random memory access} trace workload for low power target, low latency target, and jointly optimize for low power and latency. (b),(d), and (f) Mean episode return for \texttt{streaming memory} access trace workload for low power target, low latency target, and jointly optimize for low power and latency. Each RL formulation was run for five seeds each and the error bars signifies 95\% confidence intervals. PPO (MARL) is the MARL formulation. PPO (SARL), ppo (TDM), and SAC (SARL) are the single-agent formulation for the same task. Appendix~A has more information about single-agent RL formulations.\vspace{-10pt}}
\label{fig:reward-curves}
\end{figure*}

\section{Results}
This section describes our evaluation of the MARL formulation for designing a domain-specific DRAM memory controller for two different workload traces and three different optimization objectives. It also compares against single-agent RL formulations.% The two memory traces (see \Fig{fig:rand-trace} and \Fig{fig:stream-trace} in Appendix~\ref{app:workload}) are representative of real-world workloads.
 Our goal is to design a custom memory controller that achieves low power, low latency, and a joint objective of achieving low power and latency. Our results demonstrate that MARL formulations consistently outperform single-agent RL formulations across both workloads and all three optimization objectives.

% \textbf{MARL outperforms Single-Agent RL Formulations.}
We compare the MARL performance for three different design objectives: low power, low latency, and joint optimization of low power and low latency based on mean episodic return.  Given the rewards specified by Eq.~\ref{eq:r_power} are closely tied with the power, latency, and energy, a higher mean return signifies a memory controller design that achieves better architectural performance and energy efficiency.

\textit{Low Power.} \Fig{fig:rand-low-pow} and \Fig{fig:stream-low-pow} compares the performance of MARL formulation with three single-agent RL baselines for designing a low-power DRAM memory controller for random memory access trace (\Fig{fig:rand-trace}) and streaming memory access trace (\Fig{fig:stream-trace}). The reward function of all the agents is given by Eq~\ref{eq:r_power}. PPO\_TDM achieves a mean episode return of $\approx$ 4.29. Likewise, for PPO and SAC agents, the mean episode reward is 48.44 and 146 compared to MARL's mean episode reward of $\approx$ 300. Thus, MARL formulation achieves  2$\times$, 6$\times$, and 60$\times$ better mean episode return compared to PPO, SAC, and vanilla PPO RL formulations. A mean return of $\approx$ 300 corresponds to a memory controller design that dissipates 1 Watt.

\textit{Low-Latency.} \Fig{fig:rand-low-lat} and \Fig{fig:stream-low-lat} shows the performance of MARL formulation in comparison with single-agent RL formulations for designing a low latency DRAM memory controller for random and streaming memory access traces. For the low-latency objective, our evaluation shows that MARL formulation, on average, achieves $\approx$ 1.5 $\times$ to 2$\times$ more mean episodic return compared to other single-agent RL formulations.

\textit{Joint Optimization of Low Power and Low Latency.} \Fig{fig:rand-low-pow-lat} and \Fig{fig:stream-low-pow-lat} shows mean episodic return for MARL formulation with other single-agent RL formulations. MARL formulations consistently achieve $\approx$ 2-2.5$\times$ mean episodic return compared to other single-agent formulations.

Our exhaustive evaluation across five seeds across different memory access traces and optimization objectives shows that MARL outperforms single-agent RL formulations.% Next, we will understand why MARL works better than single-agent RL formulations.

\section{Discussion \blue{and Future Work}}
This work presents the first MARL formulation for the hardware architecture search problem. We show that complex interconnected systems have independent parameters that can be mapped to decentralized agents working together to achieve a target. Our exhaustive evaluation shows that our MARL approach consistently outperforms several single-agent RL formulations for hardware architecture searches for DRAM memory controllers for different workloads.

% \textbf{Future Work.} 
Despite the performance of our method, we must understand the pitfalls regarding resource constraints. While assigning one agent per parameter, as we explored in this paper, is promising, as system complexity increases this approach will also incur hardware resource overheads (each agent has its independent neural network policy). Our future work is centered around the following directions.

\textbf{Increase Computer System Complexity.} We plan to expand the scope of architecture DSE problems from DRAM systems to custom hardware accelerators or SoC designs. \blue{Increasing the complexity of the architecture DSE from DRAM to SoC designs will also result in a much larger design (million to billion times larger design space) to evalaute the efficiacy of our approach.} In addition, we want to explore how to systematically apply MARL formulations to computer architecture DSE problems as we continue to validate our MARL formulations. 

\blue{\textbf{Non-RL Baselines.} In this paper, we propose decentralized multi-agent RL  as an alternative single-agent RL for architecture DSE. In the future, we would like to benchmark the performance of our approach on other ML baselines such as Bayesian optimization~\citep{reagen2017case}, evolutionary algorithms~\citep{evo_algo}, and other bio-inspired multi-agent algorithms such as ant colony optimizations~\citep{aco}. Applying popular non-RL baselines will allow us to holistically evaluate various trade-offs in applying decentralized MARL algorithms for architecture DSE.}

\textbf{Improve MARL formulation.} We would like to systematically understand how to scale multiple agents for a given architecture DSE problem. For instance, \Fig{fig:corr_mat}, some parameters play less roles in a target objective. We believe we can use this information to prune the number of agents, thereby improving the overall performance of MARL formulation. Alternatively, we can cluster parameters belonging to the same subsystem and assign them to the same agent.

\textbf{Generalizability}. MARL, by default, provides us with a way to push the boundary on generalization. However, as we scale the complexity of the computer system, we would like to see how a group of agent's policies (let's say memory) can be reinitialized as we evaluate different workloads or a completely different computer system but having the same sub-system (e.g., memory sub-system).

%We believe answering these questions will be critical to holistically understanding how to reduce the resource overheads in applying MARL while exploiting its benefits compared to single-agent formulations.
\section{Acknowledgements}
The authors would like to thank Yanqi Zhou (Google Research, Brain Team) and Douglas Eck (Google Research, Brain Team) for reviewing this manuscript internally during the approval process. The authors would also like to thank Niko Grupen (Google Research, Brain Team) for his helpful documentation on the hyperparameter sensitivity study in ACME. The authors also thank Asma Ghandeharioun, Andrei Kapishnikov, and Yannick Assogba for making the multiagent ACME codebase available. This research is based upon work supported in part by the Google Student Researcher program and the Office of the Director of National Intelligence (ODNI), Intelligence Advanced Research Projects Activity (IARPA), via 2022-21102100013. The views and conclusions contained herein are those of the authors and should not be interpreted as necessarily representing the official policies, either expressed or implied, of ODNI, IARPA, or the U.S. Government. The U.S. Government is authorized to reproduce and distribute reprints for governmental purposes notwithstanding any copyright annotation therein.

%\section{Conclusion}
%We present a multi-agent RL for the architecture design space exploration problem. As the complexity of the computer system increases due to stringent requirements in performance and energy efficiency, we believe we also need a scalable and systematic approach to apply reinforcement learning or ML methods to quickly and intelligently navigate the large design space. Our results on domain-specific DRAM memory controller design for different workload traces and optimization constraints consistently show that MARL formulation is a promising approach compared to single-agent RL formulations.

\bibliographystyle{plainnat}
\bibliography{references}

\section*{Appendix}
\subsection{DRAM-Gym and Example Workloads}
\label{app:workload}

\begin{figure*}[b!]
  %\centering
%\begin{subfigure}[t]{0.4\columnwidth}
  %\centering
%  \includegraphics[width=0.95\columnwidth]{figs/dram-gym.png}
%  \caption{DRAM-Gym}
%  \label{fig:dram-gym}
%\end{subfigure}%
\begin{subfigure}[t]{0.48\columnwidth}
  %\centering
  \includegraphics[width=0.95\columnwidth]{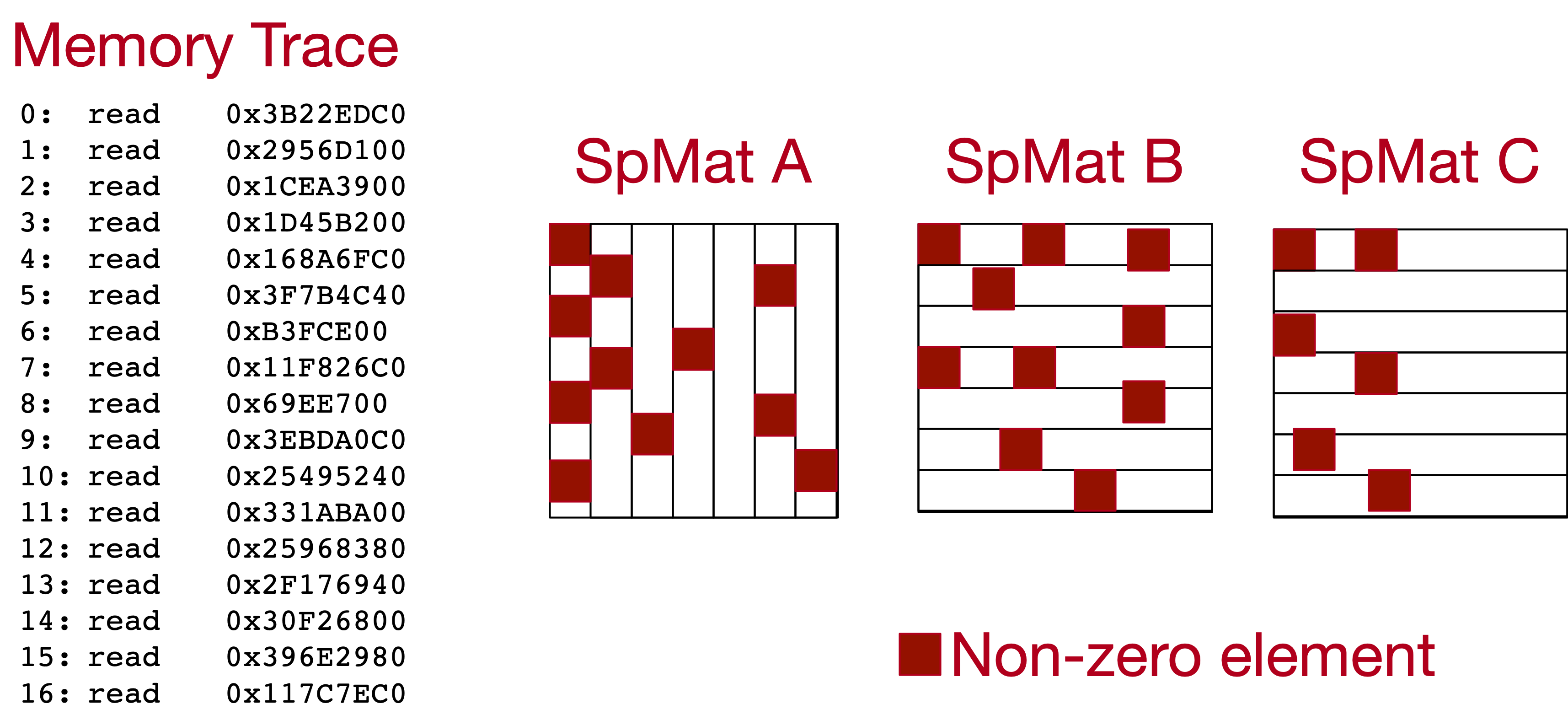}
  \caption{Random memory access trace.}
  \label{fig:rand-trace}
\end{subfigure}%
\begin{subfigure}[t]{0.48\columnwidth}
  %\centering
  \includegraphics[width=0.95\columnwidth]{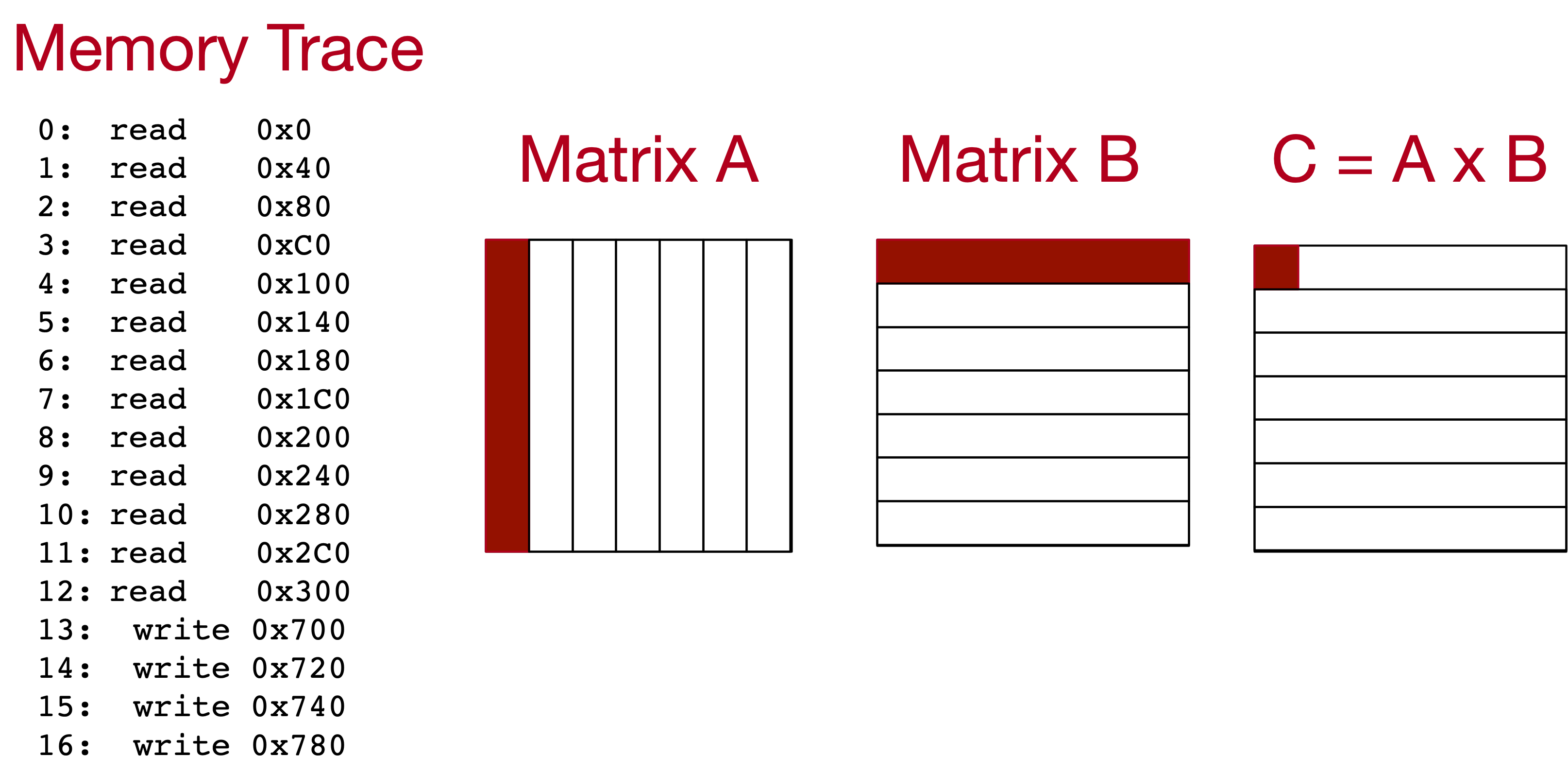}
  \caption{Streaming memory access trace.}
  \label{fig:stream-trace}
\end{subfigure}

\caption{ (a) Random memory access workload trace. Sparse matrix-matrix arithmetic results in random memory access. (c) Streaming memory access trace. Dense matrix-matrix arithmetic results in a streaming address pattern.}
\label{fig:workload_traces}
\end{figure*}

We evaluate two memory access traces: streaming access and random memory access. Both choices are representative of real-world workload. For instance, convolution (without zero-skipping) results in dense matrix arithmetic, which involves reading each element from two input matrices in a fixed stride pattern, thus resulting in a streaming access pattern, as shown in \Fig{fig:stream-trace}.

Likewise, the random memory access trace denotes system-level optimizations that exploit sparsity~\citep{sparse-nn,sparse-2}. For example, non-linear operations such as ReLU cause a lot of zero elements in the output. For latency and energy savings, multiplication with zero elements is avoided, and only the indices of the non-zero elements are stored and accessed. Since it is hard to determine the exact indices of non-zero elements in a matrix during runtime, multiplying two sparse matrices will result in a non-deterministic random access pattern, as shown in \Fig{fig:rand-trace}.

\subsection{DRAM Memory Controller Parameters}

We focus on the memory controller design within the main memory (DRAM) system. Prior work has shown that memory controller design plays a significant role in latency and energy incurred during data transfer from main memory to compute engine (e.g., cores). To model the DRAM system, we use DRAMSys~\citep{dramsys}. DRAMSys is a fast, cycle-accurate simulator that exposes DRAM architectural parameters spanning from address mapping, memory controller, memory specification (DDR3/DDR4, etc.), memory technology (HBM, DDR, etc.), and thermal properties. In this work, we focus on parameters within the memory controller and use MARL and other single-agent RL formulations to evaluate the best memory controller parameter for a given application.

The list of parameters within the memory controller is tabulated in Table~\ref{tab:mc_params}. Varying these memory controller parameters controls how to service each address request from the compute engine. Hence, the policies and parameters inside the memory controller affect the runtime latency and power for the overall application.

\label{app:params_mc}
\begin{table}[t!]
\centering
\caption{DRAMSys Memory Controller Parameters. Note that a mixture of numerical and categorical parameters are learned for the memory controller.}
\begin{tabular}{ccccc}
\toprule
\textbf{Parameter}    & \textbf{Min}     & \textbf{Max}     & \textbf{Step}            & \textbf{Type} \\
\midrule
RefreshMaxPostponed   & 1                & 8                & 1                        & Numerical     \\
RefreshMaxPulledin    & 1                & 8                & 1                        & Numerical     \\
RequestBufferSize     & 1                & 8                & 1                        & Numerical     \\
MaxActiveTransactions & 1                & 128              & $2^x$     & Numerical     \\
\midrule
\textbf{Parameter}    & \multicolumn{3}{c}{\textbf{Values}}                                      & \textbf{Type} \\
\midrule
PagePolicy            & \multicolumn{3}{c}{Open, OpenAdaptive, Closed, ClosedAdaptive} &  Categorical   \\
Scheduler             & \multicolumn{3}{c}{Fifo, FrFcfsGrp, FrFcfs}                    &  Categorical   \\
SchedulerBuffer       & \multicolumn{3}{c}{Bankwise, ReadWrite, Shared}                &   Categorical   \\
Arbiter               & \multicolumn{3}{c}{Simple, Fifo, Reorder}                      & Categorical   \\
RespQueue             & \multicolumn{3}{c}{Fifo, Reorder}                              & Categorical   \\
RefreshPolicy         & \multicolumn{3}{c}{NoRefresh, AllBank}                         & Categorical  \\
\bottomrule
\end{tabular}
\label{tab:mc_params}
\end{table}

\subsection{RL Algorithms and its Hyperparameters}
\label{sec:rl-hyper}

To remove statistical noise, we run a wide hyperparameter sweep for each evaluation involving five seeds, learning rate, and entropy values. We evaluate that trained policy at an interval of 100 training steps and report the mean episode return.

Below, we describe the hyperparameters we used for our MARL agents, TDM agents, PPO, and SAC agents, respectively.

\begin{figure*}[b!]
  \centering
\begin{subfigure}[t]{0.65\columnwidth}
  \centering
  \includegraphics[width=\columnwidth]{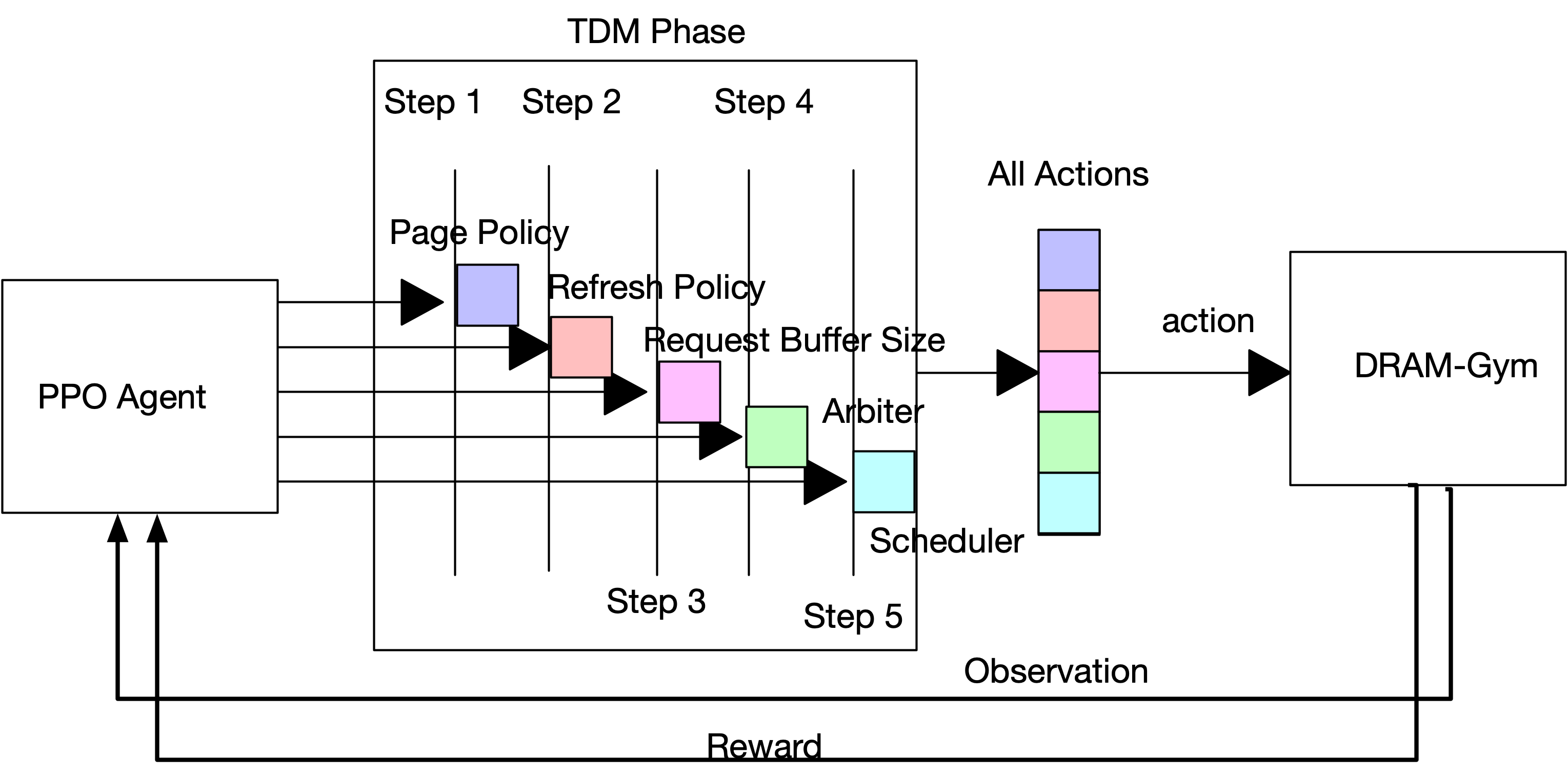}
  \caption{TDM Formulation.}
  \label{fig:tdm_form}
\end{subfigure}%
\begin{subfigure}[t]{0.4\columnwidth}%
  \centering
  \includegraphics[width=0.9\columnwidth]{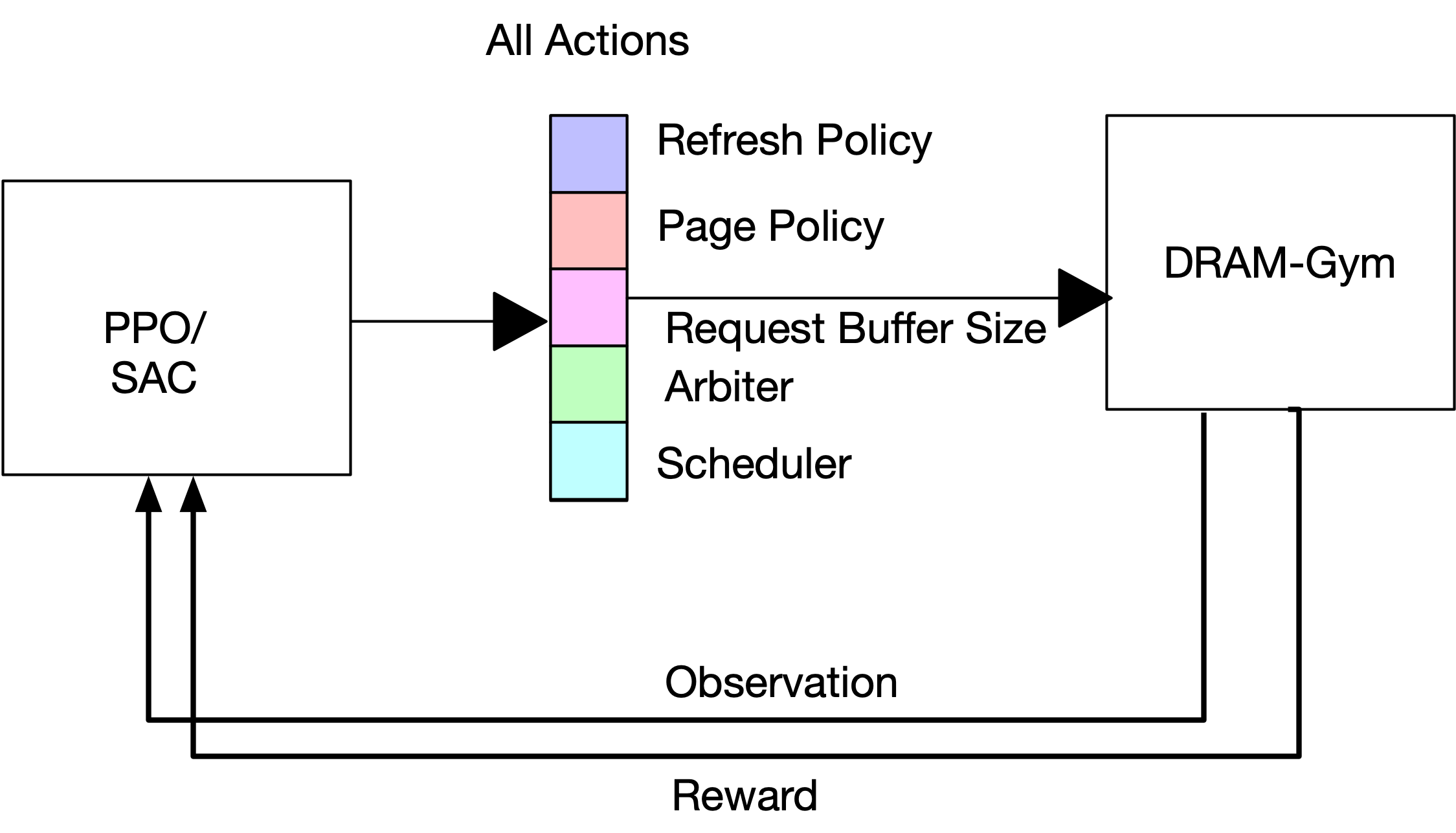}
  \caption{PPO and SAC formulations.}
  \label{fig:sa_form}
\end{subfigure}%
\caption{ (a) TDM formulation for DRAM-Gym. In TDM, at each step the agent takes action for one parameter. In TDM phase, it accumulates all the actions together and sends it to the environment. At the end of the step, it gets a non-zero reward. (b) PPO and SAC single-agent formulations.}
\end{figure*}

\begin{table}[ht]
    \centering
    \setlength{\tabcolsep}{0pt}
    \caption{The hyperparameters used in PPO and SAC agents. For the MARL formulation, each agent is an decentralized PPO agent with independent policy. Likewise for TDM, there is one PPO agent.}
    \label{tab:hyperparams}
    % \label{tab:hyperparams}
    \begin{subtable}[t]{0.48\textwidth}
        \caption{TDM/PPO/MARL Hyperparameters}
        \begin{tabular}{cc}
        \toprule
        Hyperparameter & Value \\
        \midrule
        \texttt{batch\_size}              & 128   \\
        \texttt{discount}                 & 0.99  \\
        \texttt{learning\_rate}           & \{1e-5, 2e-5, 2e-4, 1e-4\}   \\
        \texttt{adam\_epsilon}            & 1e-5  \\
        \texttt{num\_minibatch}           & 32    \\
        \texttt{unroll\_length}           & 16    \\
        \texttt{num\_epochs}              & 5     \\
        \texttt{clip\_value}              & False \\
        \texttt{clipping\_epsilon}   & 0.2   \\
        \texttt{gae\_lambda}              & 0.95  \\
        \texttt{entropy\_cost}            & 0.01  \\
        \texttt{value\_cost}              & 1.0   \\
        \texttt{max\_gradient\_norm}      & 0.5   \\
        \texttt{prefetch\_size}           & 4     \\
        \texttt{variable\_update\_period} & 1    \\
        \bottomrule
        \end{tabular}
    \end{subtable}
    \hfill
    \begin{subtable}[t]{0.48\textwidth}
        \caption{SAC Hyperparameters}
        \begin{tabular}{cc}
        \toprule
        Hyperparameter & Value \\
        \midrule
        \texttt{batch\_size}              & 256   \\
        \texttt{discount}                 & 0.99  \\
        \texttt{learning\_rate}           & \{1e-5, 2e-5, 2e-4, 1e-4\}   \\
        \texttt{reward\_scale}            & 1 \\
        \texttt{n\_step}                  & 1    \\
        \texttt{entropy\_coefficient}     & None  \\
        \texttt{target\_entropy}          & 0.0  \\
        \texttt{tau}                      & 0.005  \\
        \bottomrule
        \end{tabular}
    \end{subtable}
\end{table}

\textbf{MARL.} In our decentralized MARL formulations, we assign one agent for one memory controller parameter we control, as shown in \Fig{fig:interconnected_systems_marl}. Each agent gets the same observation. All agents take simultaneous independent actions and get a global reward (Eq~\ref{eq:r_power}). Each decentralized agent in our MARL is a PPO agent.

Table~\ref{tab:hyperparams} tabulates the hyperparameters we use for MARL agents.

\textbf{Time Division Multiplexing (TDM).} \Fig{fig:tdm_form} shows the details of the TDM formulation we use for DRAM-Gym. As an illustration, a single PPO agent at step 1 predicts the value of \texttt{Page Policy} parameter. Step 2 predicts the value for \texttt{Refresh Policy}, and so on. The number of steps in the TDM phase corresponds to the number of parameters. The agents get a zero reward for each step in the TDM phase. Once all the actions are accumulated, the actions are passed to the DRAM-gym environment. The agent gets a non-zero reward, and the next TDM iteration starts. The hyperparameters we use for the TDM phase are tabulated in Table~\ref{tab:hyperparams}.

\textbf{PPO and SAC Single Agent Fomulations.} 
\Fig{fig:sa_form} shows the single agent RL formulation we use for DRAM-Gym. We use two RL algorithms, namely PPO and SAC. At each time step, a single agent (PPO/SAC) predicts all the parameter values of the memory controller parameters. The agent receives a reward based on Eq~\ref{eq:r_power}. The hyperparameters we use for the TDM phase are tabulated in Table~\ref{tab:hyperparams}.

\end{document}